\documentclass[12pt]{iopart}
\usepackage{epsfig}

\usepackage{iopams}
\usepackage{graphicx}   % need for figures
\usepackage{color}

\newcommand{\bq}{\begin{equation}}
\newcommand{\eq}{\end{equation}}
\newcommand{\bqa}{\begin{eqnarray} \displaystyle}
\newcommand{\eqa}{\end{eqnarray}}
\newcommand{\nn}{\nonumber \\}

\begin{document}
\title[Emergence of supersymmetry on 
the surface of three dimensional 
topological insulators]{
Emergence of supersymmetry on 
the surface of three dimensional 
topological insulators}
%\date{\today}
\author{Pedro Ponte$^{1,2}$ and Sung-Sik Lee$^{1,3}$}

\address{
$^1$ Perimeter Institute for Theoretical Physics, Waterloo, Ontario, Canada N2L 2Y5}
\address{$^2$ Centro de F\'\i sica do Porto e Departamento de F\'\i sica e Astronomia, Faculdade de Ci\^encias da Universidade do Porto,
Rua do Campo Alegre 687, 4169-007 Porto, Portugal}
\address{$^3$ Department of Physics \& Astronomy, McMaster University, Hamilton, Ontario, Canada  L8S 4M1}
\ead{pponte@perimeterinstitute.ca}

\begin{abstract}
We propose two possible experimental realizations of a 2+1 dimensional spacetime supersymmetry
at a quantum critical point on the surface of three dimensional topological insulators.
The quantum critical point between the semi-metallic state with one Dirac fermion and the s-wave superconducting state on the surface
is described by a supersymmetric conformal field theory within $\epsilon$-expansion.
We predict the exact voltage dependence of the differential conductance at the supersymmetric critical point.
\end{abstract}
\submitto{\NJP}
\maketitle

%%\section{Introduction}

For the past forty years, supersymmetry has been studied intensively in high energy physics
because of its attractive features, e.g. as a possible solution to the hierarchy problem\cite{Weinberg2000}. 
Although there is so far no experimental evidence for our universe to be supersymmetric,
there is some expectation that supersymmetry may be revealed in the large hadron collider (LHC)
in a near future.
Condensed matter systems provide alternative ways to realize supersymmetry in nature through emergence\cite{SCOTT}.
Namely, supersymmetry can dynamically emerge in the low energy limit of some condensed matter systems
although the microscopic Hamiltonians explicitly break it.
Because supersymmetry is a symmetry between boson and fermion, 
it is essential to have a same number of low energy modes for boson and fermion
in order to realize supersymmetry.
Although there are examples of emergent spacetime supersymmetry in 1+1 dimensions\cite{FRIEDAN,Fendley2002,Fendley2003},
where the distinction between boson and fermion is rather obscure,
it is not easy to realize supersymmetry in lattice models in higher dimensions \cite{ROY2013}.
Because of the fermion doubling problem, 
which is actually an intrinsic feature rather 
than a `problem' in condensed matter systems, 
there are usually more fermionic degrees of freedom 
than bosonic degress of freedom
unless there is a special symmetry or dynamical mechanism
that protect multiple gapless bosonic modes at low energies\cite{Balents1998,Lee2007,Yu2010}.
On the contrary, there is no such problem in continuum model\cite{SANNINO}.

Topological insulator \cite{Konig2007,Hsieh2008,Xia2009,Hsieh2009,Chen2009,Kane2005,Bernevig2006,Fu2007a,Fu2007b,Moore2007}
 is a topological phase of matter 
where gapless edge or surface modes are protected by time reversal symmetry \cite{Xu2006,Wu2006,Andreas}.
In topological insulators, there is no fermion doubling problem 
because the second set of fermionic modes is located on the other edge or surface of a sample.
For example, on the surface of a semi infinite three dimensional topological insulator
one can have only one 2+1 dimensional Dirac fermion,
which is worth of one complex boson in terms of counting the number of propagating modes.
Therefore topological insulator provides a platform 
to realize interesting critical states\cite{Xu2010},
incluing states with emergent supersymmetry.
In this paper, we consider a superconducting quantum critical point 
on the surface of a three dimensional topological insulator.
It is likely that the critical point exhibits an emergent supersymmetry
because there are the same number of propagating modes for boson and fermion which are strongly mixed with each other at low energies.

%%\section{Emergence of supersymmetry in a pure fermionic system}

We consider a three dimensional topological insulator 
which has a gap in the bulk and a gapless Dirac fermion
at the $\Gamma$-point of the surface Brillouin zone. 
For example, Bi$_2$Se$_3$ has the desired properties\cite{Zhang2009,Hsieh2009}. 
This is an ideal material due to the large band gap in the bulk ($0.3$ eV) 
and the possibility of manipulating the Fermi energy of the bulk and the surface by chemical modifications \cite{Hasan2010}. 
We consider the case where the chemical potential is tuned to the Dirac point.
Since the dispersion relation is linear near the Dirac point, 
the low-energy excitations are described by a two-component massless Dirac fermion,
\begin{equation}
\mathcal{L}=i \bar{\psi}(\gamma_0 \partial_\tau + c_f \gamma_i \partial_i)\psi
\end{equation}
where $c_f$ is the Fermi velocity, $\gamma_0\equiv \sigma_3$, $\gamma_1\equiv \sigma_1$, $\gamma_2\equiv \sigma_2$, $\bar{\psi}\equiv -i \psi^\dagger\gamma_0$ and $\tau$ is the imaginary time.

\begin{figure}[h]
\centering
\scalebox{0.3}{\includegraphics[width = \textwidth]{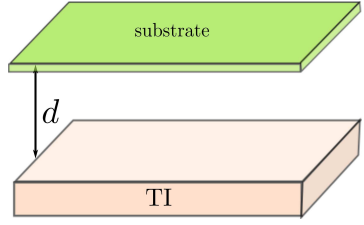}}
\caption{A metallic substrate placed above a three dimensional topological insulator.}
\label{fig:TIsubs}
\end{figure}

The two dimensional electrons on the surface are 
subject to  the electrostatic Coulomb repulsion
and the attractive interaction mediated by phonons.
If the attractive interaction is strong enough, 
the semi-metallic state can become unstable,
undergoing a quantum phase transition to 
a superconducting state.
In order to access the critical point
by tuning the strength of the Coulomb repulsion,
we consider a substrate placed at a distance $d$ above the topological insulator (Fig. \ref{fig:TIsubs}).
The substrate consists of a two-dimensional metal with dispersion relation $\epsilon=\frac{|\textbf{p}|^2}{2m}  - \mu_F$, where $\mu_F$ is the chemical potential.  
The long range Coulomb repulsion between electrons on the surface of the topological insulator 
is screened by the metallic substrate.
The strength of the residual short range repulsion can be controlled by changing $d$,
which can be used to drive the system to the critical point.
Here we assume that the attractive interaction due to phonon is sufficiently strong
so that the semi-metallic state is unstable to the s-wave superconducting state 
without the Coulomb repulsion.
Now we examine how the screened Coulomb interaction depends on 
the distance between the substrate and the topological insulator.

The system composed of the topological insulator and the substrate is described by the partition function
\begin{equation}
Z=\int D \bar{\psi} D\psi D \Psi^\dagger D\Psi Da_0 e^{-S_{TI}[\bar{\psi},\psi]-S_{s}[\Psi^\dagger ,\Psi]-S_g[a_0]-S_{int}[\bar{\psi},\psi,\Psi^\dagger ,\Psi,a_0]}\\
\label{eq:LagTIsubs}
\end{equation}
where
\begin{eqnarray}
S_{TI}&=&\int d\tau dx dy \, ~~ i \bar{\psi}(\gamma_0 \partial_\tau +c_f \gamma_1 \partial_x+c_f \gamma_2 \partial_y)\psi,\\
S_{s}&=&\sum_\sigma\int d\tau dx dy\; \Psi^\dagger_{\sigma}\left( \partial_\tau -\frac{1}{2m}(\partial^2_x+\partial^2_y) - \mu_F\right)\Psi_{\sigma},\\
S_{g}&=&
-\frac{1}{2}
\int d\tau dx dy dz ~~a_0(\partial_x^2+\partial_y^2+\partial_z^2)a_0,\\
S_{int}&=& - i \int d\tau dx dy\;
\left[ e\,\psi^\dagger \psi a_0 (\tau,x,y,0)  + \sum_\sigma e\,\Psi^\dagger_{\sigma} \Psi_{\sigma}a_0(\tau,x,y,d) \right].
\end{eqnarray}
Here $a_0$ is the temporal component of the 3+1D electromagnetic field.
We choose to work in the Coulomb gauge $\nabla\cdot \mathbf{a}=0$, 
and neglect the spatial components of the electromagnetic gauge field
whose contribution is down by $c_f/c$, where $c$ is the speed of light.
The $z$-coordinates of the surface of the topological insulator and the substrate are $0$ and $d$ respectively.
We neglect tunneling between the substrate and the topological insulator.

By integrating out the fermions of the substrate, 
we obtain the effective action for the gauge field,
\begin{equation}
 S_{0}=\frac{1}{2(2\pi)^5}\int d^3p\,dp_z\,dp_z^{\prime} a_0(p,p_z)G^{-1}(p,p_z,p_z^{\prime})a_0(-p,p_z^{\prime}),
\end{equation}
where
\begin{equation}
G^{-1}(p,p_z,p_z^{\prime})=G_0^{-1}(p,p_z)(2\pi)\delta(p_z^{\prime}+p_z)-\Pi^0(p)e^{ip_z d}e^{i p_z^{\prime}d}.
\end{equation}
Here $p=(\omega,{\bf p})$ denotes 2+1 dimensional energy-momentum vector. 
The bare propagator and the polarization (in the limit $|\textbf{p}|\ll p_F$) is given by
\bqa
 G_0^{-1}(\textbf{p},p_z)&=&|\textbf{p}|^2+p_z^2, \nn
\Pi^0(0,\textbf{p})&=&-\frac{e^2m}{\pi}.
\eqa
%%where $\textbf{p} = (p_x, p_y)$.
%\Pi^0(0,\textbf{p})&=&\frac{e^2}{4\pi\sqrt{v_x v_y}}\frac{\sqrt{\mu_F}}{\tilde{p}}\left[\left(C_+\sqrt{(z+u)^2-1}+C_-\sqrt{(z-u)^2-1}-2z\right)\right.\nonumber\\
%&&\left.+i\left(-D_{-}\sqrt{1-(z-u)^2}+D_+\sqrt{1-(z+u)^2}\right)\right].
%Here $C_{\pm}=\text{Sign}(z\pm u),\,\;D_{\pm}=0\; \text{if}\; |z\pm u|>1$ and
%$C_{\pm}=0,\,\;D_{\pm}=1\; \text{if}\; |z\pm u|<1$.
%$|\textbf{p}| = \sqrt{p_x^2 + p_y^2}$, $\tilde{p}=\sqrt{v_x p_x^2+v_y p_y^2}$, $z=\frac{p}{2\sqrt{\mu}}$ and $u=\frac{\omega}{2p\sqrt{\mu}}$.\cite{PhysRevLett.18.546}
It is noted that the presence of the substrate breaks the translation invariance along the $z$-direction,
and the momentum along this direction is not conserved. 
Inverting the dressed propagator, we obtain the two dimensional screened Coulomb repulsion between electrons on the surface of the topological insulator,
\bqa
 V(p) & =& \int \frac{dp_z}{2\pi} \frac{dp_z^{'}}{2\pi} G(p,p_z,p_z^{\prime}) \nn
& = & \frac{e^2}{4 |\textbf{p}|}\left(1+\frac{\Pi^0(p)e^{-2|\textbf{p}|d}}{2|\textbf{p}|-\Pi^0(p)}\right).
\eqa
%where $|\textbf{p}| = \sqrt{p_x^2 + p_y^2}$.
The static effective potential is not singular as $|\textbf{p}|\rightarrow 0$. 
%In the limit $u\ll 1$, the effective potential is not singular as $|\textbf{p}|\rightarrow 0$. 
It is given by $V_0=\frac{e^2}{2}\left(  d + \frac{1}{e^2N(\mu_F)}\right)$, 
where $N(\mu_F)$ is the density of states of the substrate at the Fermi energy. 

If $N(\mu_F)$ is large and $d$ is small,
the Coulomb repulsion can be made weak so that
the attractive interaction mediated by phonon dominates.
If the strength of the attractive interaction is sufficiently strong,
one can tune the system across the 
superconducting phase transition 
by changing $d$.

In the presence of one Dirac point located at the $\Gamma$ point, one can, in general, have a pairing of the form $\Delta_{\alpha,\beta}(\textbf{k}) \psi_{\alpha,\textbf{k}} \psi_{\beta, -\textbf{k}}$, where $\alpha,\beta=1,2$ are pseudospin indices and  
 $\Delta_{\alpha,\beta}(\textbf{k}) = -\Delta_{\beta, \alpha}(-\textbf{k})$.  The gap function can be decomposed as 
 $\Delta_{\alpha, \beta}(\textbf{k}) =  \epsilon_{\alpha, \beta} \Delta_s(\textbf{k})  + [ \sigma_y  \vec \sigma ]_{\alpha,\beta} \cdot \vec \Delta_t(\textbf{k}) $, where $\Delta_s(\textbf{k})$ and $\vec \Delta_t(\textbf{k})$ are pseudospin singlet and triplet order parameters, respectively.
It is expected that the triplet state is energetically less favourable than the singlet state 
because the gap vanishes at $\textbf{k}=0$ for the triplet state.  
However, this ultimately depends on the microscopic details of the systems. Here we proceed with the assumption that the s-wave singlet superconducting state is the dominant instability channel
in the presence of the strong attractive interaction. 

Suppose that electrons are interacting through a net attractive interaction
\begin{equation}
U=-v_0 \psi^\dagger \psi \psi^\dagger \psi. 
\label{eq:attractive}
\end{equation}
To decouple the four fermion interaction we introduce a complex boson field through the Hubbard-Stratonovich transformation,
\begin{equation}
-\frac{v_0}{2} (\psi^T \varepsilon \psi)^\dagger (\psi^T \varepsilon \psi)\rightarrow  (\psi^T \varepsilon \psi)^\dagger \phi+\phi^*(\psi^T \varepsilon \psi)+\frac{2|\phi|^2}{v_0}.
\label{eq:HS}
\end{equation}
Here $\varepsilon$ is the $2\times 2$ antisymmetric matrix with $\varepsilon_{12}=-\varepsilon_{21}=1$.
Although the complex boson (Cooper pair) has no bare kinetic term, 
it is generated by fermions at low energies. 
The low energy effective theory becomes 
\begin{eqnarray}
\mathcal{L}&=&i \bar{\psi}(\gamma_0 \partial_\tau +c_f \gamma_i \partial_i)\psi
+|\partial_\tau \phi|^2+c_b^2|\partial_i\phi|^2+m^2|\phi|^2+ \lambda |\phi|^4\nn
&&+ h \left(\phi^*\psi^T \varepsilon \psi- \phi\bar{\psi} \varepsilon \bar{\psi}^{T}\right), 
\label{eq:LSUSY}
\end{eqnarray}
where $c_b$ is the velocity of bosons
which may be different from $c_f$.
Note that the dynamics of bosons is guaranteed to be relativistic with the dynamical critical exponent $z=1$
as far as fermions are relativistic
because Cooper pairs are formed out of the relativistic Dirac fermions.

As $d$ is tuned, the mass of the boson is changed.
In the superconducting state with $m^2 < 0$, 
the boson is condensed, and the Dirac fermion becomes gapped.
At a critical distance $d_c$, bosons are massless
and the theory flows to an interacting fixed point in the low energy limit.
The field theory which has two copies of the present theory has been studied,
where each set of modes describes one Dirac fermion and one complex boson 
defined at one of the two distinct momentum points ($K$ and $K^{'}$) 
on the honeycomb lattice\cite{Lee2007}.
In the $\epsilon$ expansion, it has been shown that the theory flows to 
a supersymmetric critical point with four emergent supercharges
where two sets of modes are decoupled in the low energy limit.
Therefore the same conclusion can be drawn for the present case.
In the low energy limit, the critical point is described by 
the ${\cal N}=2$ Wess-Zumino theory with one chiral multiplet \cite{Wess1974}.

%%\subsection{Proximity effect between a Josephson Junction Array and a strong topological insulator}\label{sec:JJA_STI}

\begin{figure}[h]
\centering
\scalebox{0.4}{\includegraphics[width = \textwidth]{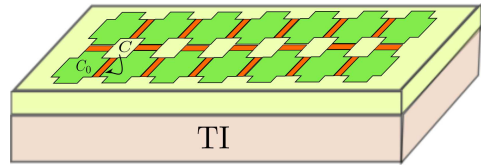}}
\caption{A Josephson junction array on the surface of a three dimensional topological insulator. The capacitances $C_0$ and $C$ controls the charging energy and the tunneling of Cooper pairs respectively.}
\label{fig:JJATI}
\end{figure} 

In order for an intrinsic superconducting state to be stable,
one needs to have a sufficiently strong electron-phonon interaction,
which may or may not be the case for real materials.
In cold atom systems, strength of interaction can be easily tuned.
It will be of interest to realize 3D topological insulators and the supersymmetric quantum critical point in cold atom systems by tuning attractive interaction between particles\cite{DELGADO,SPIELMAN}.
Here we propose a second scenario which realizes supersymmetry in the low energy limit. 
The system consists of a Josephson junction array 
deposited on the top of a topological insulator (Fig. \ref{fig:JJATI}). 
A Josephson junction array (JJA) consists of a regular network of superconducting islands coupled by tunnel junctions. 
%Cooper pairs can hop between nearest neighbor islands and the hopping term $t$ is proportional to the Josephson energy $E_J$. By applying an external gate voltage, it is possible to tune the chemical potential. In the self-charging limit $C_0\gg C$, the interaction between Cooper pairs becomes short-ranged \cite{Fazio,PhysRevB.48.3316,PhysRevB.43.5307}.
The Dirac electrons of the topological insulator can tunnel to the JJA to form Cooper pairs and vice versa. 
The JJA and the Dirac fermion are also described by the same  Lagrangian in Eq. \ref{eq:LSUSY} at low energies
if the average number of Cooper pairs within each island is tuned to be an integer.
As the Josephson coupling is tuned, the JJA can undergo a phase transition to a superconducting state.
The quantum critical point is again described by the ${\cal N}=2$ Wess-Zumino theory with one chiral multiplet.

%%\section{Physical consequences}

At the critical point, the scaling dimensions of the chiral primary fields,
including the Dirac fermion and the Cooper pair field, 
are constrained by the superconformal algebra\cite{MINWALLA}.
As a result, the exact anomalous dimensions of the fermion and boson 
are given by $\eta_\phi=\eta_\psi=\frac{1}{3}$ \cite{Aharony1997}. 
It is of interest to provide a clear experimental signature for the emergent supersymmetry.
Here we consider tunneling spectroscopy which measures the local density of states.
%%The differential conductance $\frac{dI}{dV}$ measures the density of states \cite{bruus2004many}. 
%%finite part of the diagram (a) in (Fig. \ref{fig:oneloop}) gives the following contribution to the self-energy of the electron
% The one-loop fermion self energy is given by 
% \begin{equation}
% 	\Sigma(k)=(-k \cdot \boldsymbol \sigma )\frac{4 h^{2}}{(4\pi)^2}\left( 1+\frac{\gamma-\log(4\pi)}{2}+\frac{1}{2}\log(k^2/\mu^2)\right),
% 	\label{eq:selfenergy}
% \end{equation}
% where $\mu$ is the energy scale at which the system is probed and $\gamma$ is Euler's constant.
%According to Dyson's equation $\mathcal{G}^{-1}=\mathcal{G}^{-1}_0-\Sigma$, 
The single particle Green's function of electron at the supersymmetric critical point is given by
\begin{equation}
	\mathcal{G}(p)\sim \frac{(p\cdot \sigma)}{(p^2)^{5/6}},
\end{equation}
where $\sigma=(\sigma_3,c_f\,\sigma_1,c_f\sigma_2)$.
Integrating the spectral function over momentum, 
we obtain  that the local density of states
\begin{equation}
\rho(\omega)\sim|\omega|^{4/3}. 
\end{equation}
As a result, we expect the differential conductance $\frac{dI}{dV} \sim V^{4/3}$ at the supersymmetric critical point.
This provides a clear experimental signature of the supersymmetric state.
Note that the exact exponent is predicted thanks to the superconformal symmetry,  
although the critical point is described by the strongly interacting theory.

%%\section{Conclusion}
In summary, we propose that a 2+1D superconformal field theory can be realized at the quantum critical point between the semi-metallic state and the s-wave superconducting state on the surface of three dimensional topological insulators.
The critical point is described by the $\mathcal{N}=2$ Wess-Zumino model with one chiral multiplet.
At the critical point, the local density of states obeys the power law
behavior $\rho(\omega)\sim \omega^{4/3}$,
which can be measured by scanning tunneling microscopes.
Recently, 
it has been shown that the supersymmetric critical point is not stable in the presence of disorder\cite{NANDKISHORE}. 
However,  the critical behaviour governed by the putative supersymmetric critical point can be observed within a finite temperature range in the weak disorder limit.

\ack
We would like to thank
Patrick Lee 
for helpful discussions.
This work is a result of the research essay performed 
as a part of the Perimeter Scholars International (PSI) program.
This research was supported in part by NSERC and ERA.
Research at the Perimeter Institute is supported 
in part by the Government of Canada 
through Industry Canada, 
and by the Province of Ontario through the
Ministry of Research and Information.
\vspace{0.4 cm}

After the completion of the draft, 
we became aware of a recent preprint\cite{Grover2012} 
by Tarun Grover and Ashvin Vishwanath,
who reached the same conclusion by studying 
various instabilities on the surface of topological insulators 
in a more general context.
We thank the authors for their helpful feedbacks on our draft.

\section*{References}
%\bibliography{bibl.bib}

\begin{thebibliography}{10}

\bibitem{Weinberg2000}
For example, see Weinberg S 2000 {\it The Quantum Theory of Fields. Vol. 3: Supersymmetry}

\bibitem{SCOTT} Thomas S 2005 {\it Emergent Supersymmetry} KITP talk 

\bibitem{FRIEDAN} Friedan D, Qui Z and Shenkar S 1985 {\it Phys. Lett. B} {\bf 151} 37

\bibitem{Fendley2002}
Fendley P, Schoutens K and Boer J  2003 {\it Phys. Rev. Lett.}  {\bf 90} 120402

\bibitem{Fendley2003}
Fendley P, Nienhuis B and Schoutens K 2003 {\it Journal of Physics A} {\bf 36} 12399

\bibitem{ROY2013}
Roy B, Juricic V and Herbut I 2013 {\it Phys. Rev. B} {\bf 87} 041401
\bibitem{Balents1998}
Balents L, Fisher MPA, and Nayak C 1998 {\it International Journal of Modern Physics B} {\bf 12}

\bibitem{Lee2007}
Lee S.-S. 2007 {\it Phys. Rev. B} {\bf 76} 075103

\bibitem{Yu2010}
Yu Y and Yang K 2010 {\it Phys. Rev. Lett.} {\bf 105} 150605



\bibitem{SANNINO}
Antipin O, Mojaza M, Pica C and Sannino F 2011 Magnetic Fixed Points and Emergent Supersymmetry {\it Preprint} arXiv:1105.1510




\bibitem{Kane2005}
Kane C L and Mele E J 2005 {\it Phys. Rev. Lett.} {\bf 95} 146802


\bibitem{Bernevig2006}
Bernevig B A, Hughes T L  and Zhang S C 2006 {\it Science} {\bf 314} 1757

\bibitem{Fu2007a}
Fu L and Kane C L 2007 {\it Phys. Rev. B} {\bf 76} 045302

\bibitem{Fu2007b}
Fu L, Kane C L and Mele E J 2007 {\it Phys. Rev. Lett.} {\bf 98} 106803

\bibitem{Moore2007}
Moore J E and Balents L 2007 {\it Phys. Rev. B}  {\bf 75} 121306

\bibitem{Konig2007}
K{\"o}nig M, Wiedmann S, Br{\"u}ne C, Roth A, Buhmann H, Molenkamp LW,
 Qi X L and Zhang S C 2007 {\it Science} {\bf 318} 766

\bibitem{Hsieh2008}
Hsieh D, Qian D, Wray L,  Xia Y Q, Hor Y S, Cava R J and Hasan M Z 2008 {\it Nature} {\bf 452} 970 

\bibitem{Xia2009}
Xia Y, Qian D, Hsieh D, Wray L, Pal A, Lin H, Bansil A, Grauer D,
Hor Y S, Cava R J, et~al 2009 {\it Nature Physics} {\bf 5} 398

\bibitem{Hsieh2009}
Hsieh D, Xia Y, Qian D, Wray L, Dil J H, Meier F, Osterwalder J,
  Patthey L, Checkelsky J G, Ong N P, et~al 2009 {\it Nature} {\bf 460} 1101

\bibitem{Chen2009}
Chen Y L, Analytis J G, Chu J H, Liu Z K, Mo S K, Qi X L, Zhang H J, Lu D H,
  Dai X, Fang Z, et~al 2009 {\it Science} {\bf 325} 178 

\bibitem{Xu2006}
Xu C and Moore J E 2006 {\it Phys. Rev. B} {\bf 73}  045322 


\bibitem{Wu2006}
Wu C, Bernevig B A and Zhang S C 2006 {\it Phys. Rev. Lett.} {\bf 96} 106401

\bibitem{Andreas} Schnyder A P, Ryu S, Furusaki A and  Ludwig A W W 2008 {\it Phys. Rev. B} {\bf 78} 195125

\bibitem{Xu2010} Xu C 2010 {\it Phys. Rev. B} {\bf 81} 020411

\bibitem{Zhang2009}
Zhang H, Liu C X, Qi X L, Dai X, Fang Z and Zhang S C 2009 {\it Nature Physics} {\bf 5} 438

\bibitem{Hasan2010}
Hasan M Z and Kane C L 2010 {\it Rev. Mod. Phys.}  {\bf 82} 3045

\bibitem{Wess1974}
Wess J and Zumino B 1974 {\it Nucl. Phys. B} {\bf 70} 39 

\bibitem{MINWALLA} Minwalla S 1998 {\it Adv. Theor. Math. Phys.} {\bf 2} 781


\bibitem{Aharony1997}
Aharony O, Hanany A, Intriligator K, Seiberg N and Strassler M J 1997 {\it Nucl. Phys. B} {\bf 499} 67 


\bibitem{DELGADO}
Bermudez A, Mazza L, Rizzi M, Goldman N, Lewenstein M and
 Martin-Delgado M A 2010
{\it Phys. Rev. Lett.} {\bf 105} 190404

\bibitem{SPIELMAN}
Lin Y J, Jimenez-Garcia K and Spielman I B 2011
{\it Nature} {\bf 471} 83

\bibitem{NANDKISHORE}
Nandkishore R, Maciejko J, Huse D A and Sondhi S L 2013 {\it Phys. Rev. B} {\bf 87}  174511

\bibitem{Grover2012}
Grover T and Vishwanath A 2012 Quantum Criticality in Topological Insulators and Superconductors: Emergence of Strongly Coupled Majoranas and Supersymmetry  {\it Preprint} arXiv:1206.1332

\end{thebibliography}
%\bibliographystyle{apsrev}

\end{document}